\documentclass[preprint]{aastex}
\usepackage{natbib}
\usepackage{booktabs}
\usepackage{threeparttable}

\usepackage{longtable}
\usepackage{tabularx}
\usepackage{lineno}

\begin{document}

\title{Detection of flare multi-periodic pulsations in mid-ultraviolet Balmer continuum, Ly$\alpha$, hard X-ray, and radio emissions simultaneously}

\author{Dong~Li\altaffilmark{1,2}, Mingyu~Ge\altaffilmark{3}, Marie~Dominique\altaffilmark{4}, Haisheng~Zhao\altaffilmark{3}, Gang~Li\altaffilmark{3}, Xiaobo~Li\altaffilmark{3}, Shuangnan~Zhang\altaffilmark{3,5}, Fangjun~Lu\altaffilmark{3}, Weiqun~Gan\altaffilmark{1,6}, and Zongjun~Ning\altaffilmark{1,6}}
\affil{$^1$Key Laboratory of Dark Matter and Space Astronomy, Purple Mountain Observatory, CAS, Nanjing 210023, PR China \\
     $^2$State Key Laboratory of Space Weather, Chinese Academy of Sciences, Beijing 100190, PR China \\
     $^3$Key Laboratory of Particle Astrophysics, Institute of High Energy Physics, Chinese Academy of Sciences, Beijing 100049, PR China \\
     $^4$Solar-Terrestrial Centre of Excellence/SIDC, Royal Observatory of Belgium, 3 Avenue Circulaire, B-1180 Uccle, Belgium \\
     $^5$University of Chinese Academy of Sciences, Chinese Academy of Sciences, Beijing 100049, PR China \\
     $^6$School of Astronomy and Space Science, University of Science and Technology of China, Hefei 230026, PR China}
     \altaffiltext{}{Correspondence should be sent to: lidong@pmo.ac.cn}
\begin{abstract}
Quasi-periodic pulsations (QPPs), which usually appear as temporal
pulsations of the total flux, are frequently detected in the light
curves of solar/stellar flares. In this study, we present the
investigation of non-stationary QPPs with multiple periods during
the impulsive phase of a powerful flare on 2017 September 06, which
were simultaneously measured by the Large-Yield RAdiometer (LYRA)
and the Hard X-ray Modulation Telescope (Insight-HXMT), as well as
the ground-based BLENSW. The multiple periods, detected by applying
a wavelet transform and Lomb-Scargle periodogram to the detrended
light curves, are found to be $\sim$20$-$55~s in the Ly$\alpha$ and
mid-ultraviolet Balmer continuum emissions during the flare
impulsive phase. Similar QPPs with multiple periods are also found
in the hard X-ray emission and low-frequency radio emission. Our
observations suggest that the flare QPPs could be related to
nonthermal electrons accelerated by the repeated energy release
process, i.e., triggering of repetitive magnetic reconnection, while
the multiple periods might be modulated by the sausage oscillation
of hot plasma loops. For the multi-periodic pulsations, other
generation mechanisms could not be completely ruled out.
\end{abstract}


\section{Introduction}
Solar flares are powerful eruption events on the Sun associated with
a rapid and violent release of magnetic free energy through a
reconnection process. A typical flare can radiate at almost all
wavelengths constituting the solar spectrum, ranging from radio
through optical and ultraviolet (UV) to soft/hard X-ray (SXR/HXR)
and even $\gamma$-rays \citep[e.g.,][]{Benz17,Tan20}. Only a small
part of the flare radiation is emitted at the shortest wavelengths
in the X-ray and extreme UV (EUV) ranges \citep{Emslie12}. The
quantitative estimation of the radiated flare energy partition
suggested that about 70\% in white light (WH) for solar flares
\citep[e.g.,][]{Kretzschmar11} and 55\%$-$80\% in WL for stellar
flares \citep[e.g.,][]{Kuznetsov21}. In other words, most of the
flare energy is radiated in the longer wavelengths \citep{Kleint16}.
Between those extremes, the solar UV spectrum from 1000~{\AA} to
3000~{\AA}, which can be further split into the far-ultraviolet
(FUV), the mid-ultraviolet (MUV), and the near-ultraviolet (NUV), is
thought to provide an important contribution to the flare radiation
\citep{Woods06,Milligan14,Dominique18}. For instance, the Ly$\alpha$
spectral line produced by the chromospheric neutral hydrogen, which
is centered at 1216~{\AA} (in the FUV spectrum), is among the
spectral lines in which flares radiate the most
\citep{Allred05,Curdt01,Lu21}. The hydrogen Balmer continuum emitted
during flares, which is thought to be generated during the
recombination of flare-produced free electrons in the chromosphere,
is often detected in the MUV and NUV ranges, as well as close to the
Balmer recombination edge at 3646~{\AA}
\citep{Heinzel14,Kotrc16,Dominique18}. Both the Ly$\alpha$ and
hydrogen Balmer continuum emissions during solar flares are expected
to be nonthermal profiles, i.e., similar to the HXR radiation which
is produced by the beam of electrons that are accelerated by the
magnetic reconnection during the solar flare
\citep[e.g.,][]{Avrett86,Rubio09,Heinzel14}.

Quasi-periodic pulsations (QPPs) often refer to the quasi-periodic
intensity variations during solar/stellar flares \citep[see][for a
recent review]{Zimovets21}. In many observations, the flare QPPs
were found to show a non-stationary property in the time series
integrated over the whole Sun/star or over the oscillation region,
for instance, each pulsation has an anharmonic and symmetric
triangular profile shape \citep[e.g.,][]{Kolotkov15,Nakariakov19}.
The signature of flare QPPs can be detected in flare light curves
across a broad band of the electromagnetic spectrum, i.e.,
radio/microwave emissions
\citep{Ning05,Reznikova11,Nakariakov18,Yu19}, UV/EUV wavelengths
\citep{Shen18,Hayes19,Reeves20,Miao21}, SXR/HXR and $\gamma$-rays
channels \citep{Nakariakov10,Ning17,Hayes20,Li20a}, as well as the
H$\alpha$ \citep{Srivastava08,Kashapova20,Li20b} or Ly$\alpha$
\citep{Van11,Milligan17,Li21} emissions. The quasi-periods of these
QPPs were reported from sub-seconds to tens of minutes
\citep[e.g.,][]{Tan10,Shen13,Shen19,Kolotkov18,Karlicky20,Clarke21}.
It should be stated that the observed periods are generally related
to the specific channels or flare phases
\citep{Tian16,Dennis17,Pugh19}, suggesting that the various classes
of QPPs could be produced by different generation mechanisms
\citep[e.g.,][]{Kupriyanova20}. In the literature, the flare-related
QPPs were most often explained by magnetohydrodynamic (MHD) waves,
more specifically, sausage waves, kink waves, and slow waves
\citep{Li20c,Nakariakov20,Wang21}, or by a repetitive regime of
magnetic reconnection that could be spontaneous (i.e.,
self-oscillatory process) or triggered due to external MHD
oscillations \citep{Thurgood17,Yuan19,Clarke21}. They can also be
interpreted in terms of the LRC-circuit oscillation in
current-carrying loops \citep{Tan16,Li20b} or caused by the
interaction between supra-arcade downflows and flare loops
\citep{Xue20,Samanta21}.

The hydrogen Balmer continuum enhancement at MUV wavelengths around
2000~{\AA} was found to be highly synchronous with the enhancement
of Ly$\alpha$ emission during a powerful solar flare
\citep{Dominique18}, while the flare radiation in the Ly$\alpha$ and
HXR ranges were demonstrated to have a close relationship
\citep{Nusinov06,Jing20,Lu21b}. However, flare-related QPPs were
rarely observed simultaneously in these channels. In this study, we
report the detection of flare-related QPPs with multi-periodicity in
the MUV Balmer continuum, Ly$\alpha$, HXR and radio emissions during
the impulsive phase of a powerful solar flare.

\section{Observations and Instruments}
\begin{figure}
\centering
\includegraphics[width=\linewidth,clip=]{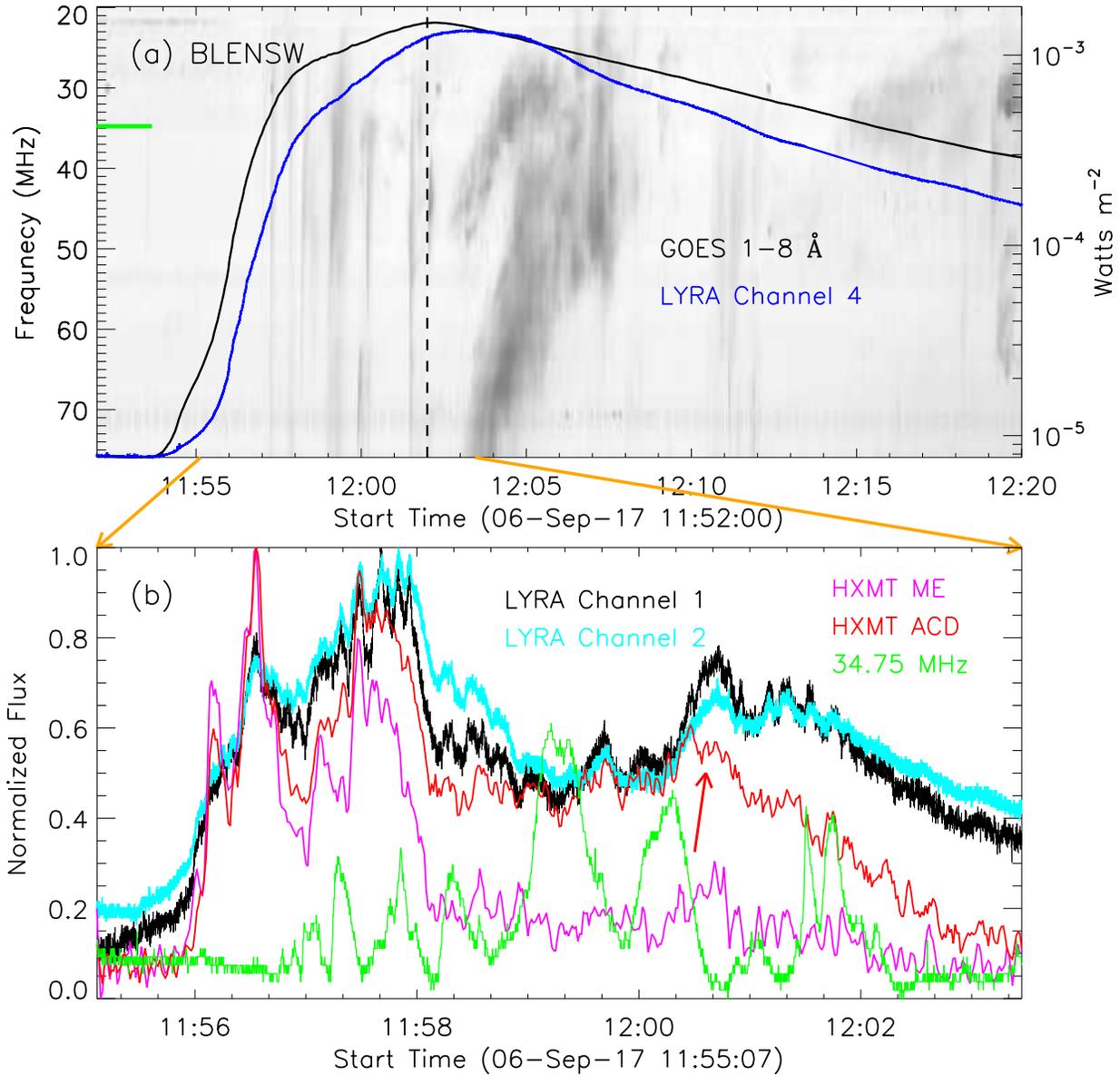}
\caption{Light curves integrated over the whole Sun during the X9.3
flare on 2017 September 06, recorded by the GOES, the LYRA, the
Insight-HXMT, and the BLENSW, respectively. Notice that, all the
light curves expect for the GOES flux are shown as normalization.
The context image in panel (a) is the radio dynamic spectra measured
by the BLENSW. The short green line marks the radio frequency at
34.75~MHz, which is plotted in panel~(b), the dashed line indicates
the flare peak time. \label{flux}}
\end{figure}

On 2017 September 06, the active region of NOAA~12673 produced the
most powerful flare of the solar cycle 24, measured to be of the
X9.3 class. It was simultaneously recorded by the space-based
instruments of Large-Yield RAdiometer (LYRA) on board the PRoject
for OnBoard Autonomy 2 (PROBA2) mission \citep{Dominique13}, the
Hard X-ray Modulation Telescope (Insight-HXMT) \citep{Zhang20}, and
the Geostationary Operational Environmental Satellite 16 (GOES-16),
as well as the ground-based CALLISTO radio spectrograph
\citep{Benz09} at BLENSW, as shown in Figure~\ref{flux} and
table~\ref{tab1}. Note that all the space- and ground-based
instruments observe in a Sun-as-a-star mode. The GOES SXR flux at
1$-$8~{\AA} (black curve) suggests that the powerful flare begins at
about 11:53~UT and reaches its maximum at around
12:02~UT\footnote{https://www.solarmonitor.org/?date=20170906}, as
indicated by the vertical dashed line in Figure~\ref{flux}~(a).

LYRA provides the solar irradiance measurement in four wide spectral
channels with a high time resolution of 0.05~s
\citep{Dominique13,Dominique18}. Channels~3 and 4 measure the solar
radiation in SXR/EUV ranges at 1$-$800~{\AA} and 1$-$200~{\AA}, and
they look very similar. Therefore, only the light curve from
channel~4 (blue) is shown in Figure~\ref{flux} (a), which exhibits a
similar time evolution as the GOES SXR flux. However, the flux peak
is a bit later than that of the GOES SXR flux. This could be
attributed to the observational fact that the light curve from LYRA
channel~4 measures the solar SXR/EUV radiation in a long and broad
wavelength range, i. e., 1$-$200~{\AA} \citep{Dominique18}, while
the GOES SXR flux only covers a short and narrow wavelength range of
1$-$8~{\AA}. On the other hand, the channel~1 observes the solar
irradiance in the FUV channel centered at the Ly$\alpha$~1216~{\AA}
line, which is also referred to as the Ly$\alpha$ channel. While the
channel~2 takes the solar observation in the MUV channel between
1900~{\AA} and 2220~{\AA}, which is demonstrated to be consistent
with the hydrogen Balmer continuum emission around 2000~{\AA}, and
is formed in the chromosphere \citep[e.g.,][]{Dominique18}.
Therefore, the LYRA channel~2 is considered to provide the solar
radiation in the MUV Balmer continuum. Figure~\ref{flux}~(b) draws
the normalized light curves between 11:55:07~UT and 12:03:27~UT from
LYRA channels in the Ly$\alpha$ (black) and MUV Balmer continuum
(cyan) ranges, showing a well synchronous relationship in the time
series. Taking into account that LYRA observes in a Sun-as-a-star
mode, it is impossible to conclude that they are radiated from a
same source area, due to the lack of the spatially resolved
information.

In this study, two X-ray light curves measured by the Insight-HXMT
\citep{Zhang20}, which has a time cadence of 1~s, were also used to
investigate the flare-related QPPs. The Medium Energy X-ray
telescope (ME) observes the X-ray emission normally at 5$-$30~keV
\citep{Cao20,Luo20}. The Anti-Coincidence Detectors (ACD) provides
the HXR flux at higher energy, i.e., $>$100~keV, which was adopted
from the High Energy X-ray telescope after removing the background
that is induced by the particle \citep{Liu20}. Figure~\ref{flux}~(b)
presents the X-ray light curves as normalization during
11:55:07$-$12:03:27~UT derived from the ME (magenta) and ACD (red).
They both match well with the LYRA light curves at channels~1 and 2,
but the ACD light curve seems to show a much clear enhancement after
12:00~UT, as indicated by the red arrow. Moreover, there is almost
no time delay between the light curves measured by the LYRA and the
Insight-HXMT, both of which have very burst profiles, confirming
their nonthermal profiles.

The major X9.3 flare was also observed by the radio spectrogram from
the BLENSW \citep{Benz09} at low frequencies between about 20~MHz
and 76~MHz, which has a time cadence of 0.25~s, as shown by the
context image in Figure~\ref{flux}~(a) and table~\ref{tab1}. A
sequence of transient bursts can be found in the radio dynamic
spectrum during the flare impulsive phase, i, e., during
$\sim$11:57$-$12:02~UT. They all drift rapidly from high to low
frequencies over a quite short time, which can be regarded as the
type III radio bursts and could be helpful to trace the propagating
electron beams through the solar atmosphere during solar flares.
Figure~\ref{flux}~(b) also draws the normalized radio flux (green)
at a low frequency of 34.75~MHz, which has been shifted in height to
show clearly. It exhibits a burst profile, which is similar to the
light curves recorded by the LYRA and the Insight-HXMT. Obviously,
the onset time of the enhancement in radio flux at the low frequency
is later than that in light curves at Ly$\alpha$, MUV Balmer
continuum and HXR emissions.

\section{Results}
\begin{figure}
\centering
\includegraphics[width=\linewidth,clip=]{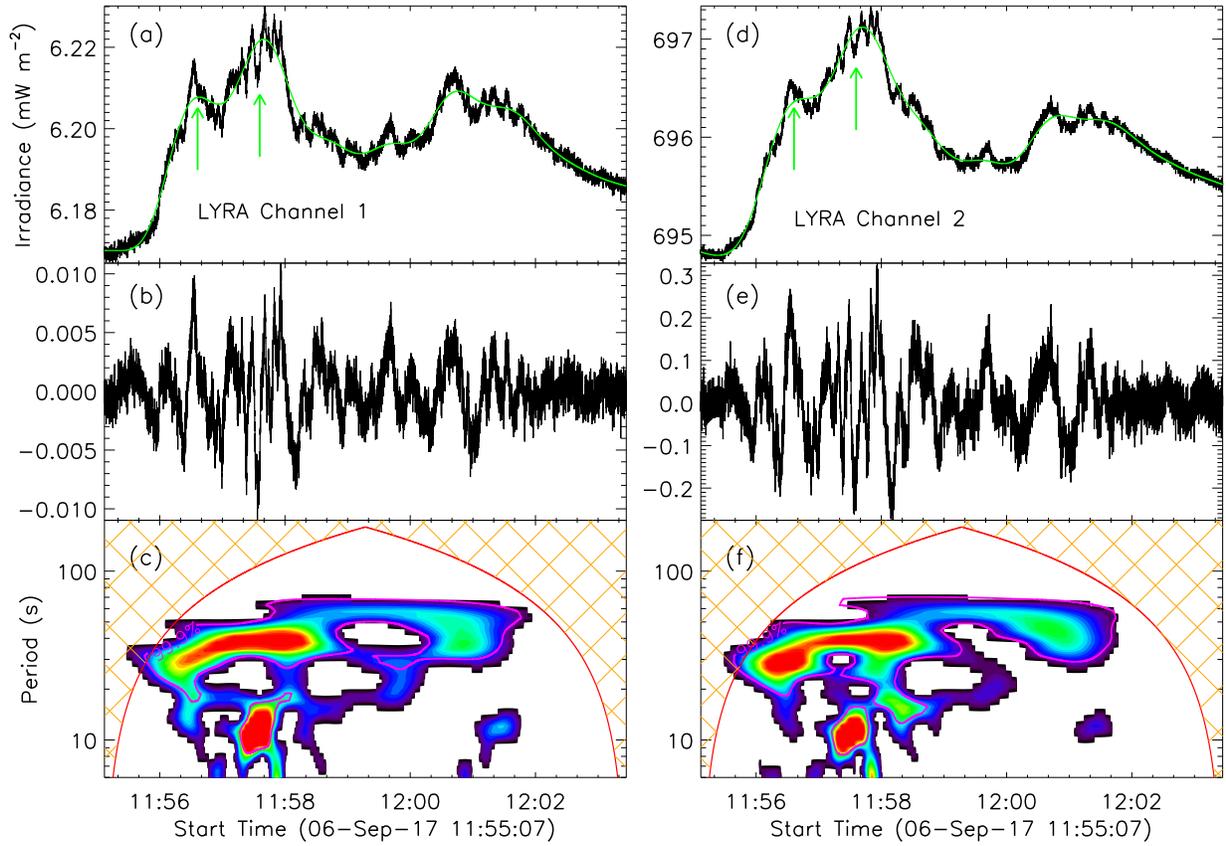}
\caption{Panels~(a) \& (d): Full-disk light curves (black) observed
by LYRA at channels~1 and 2, and their FFT filtered profiles
(green). Two green arrows marked the double main peaks in the raw
light curves. Panels~(b) \& (e): detrended light curves. Panels~(c)
\& (f): Morlet wavelet power spectra. The magenta lines outline the
significance level of 99.9\%. \label{wv_l}}
\end{figure}

\begin{figure}
\centering
\includegraphics[width=\linewidth,clip=]{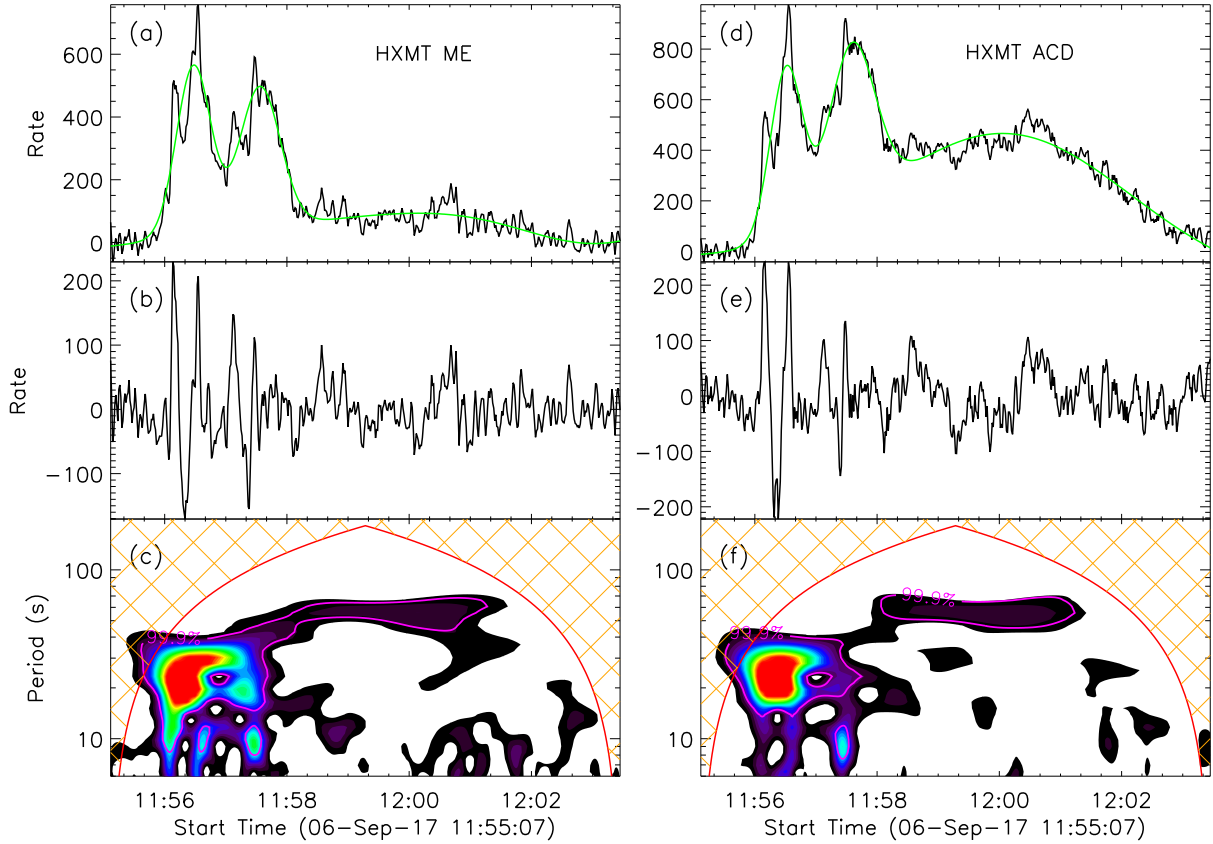}
\caption{Panel~(a), (d): Full-disk light curves (black) observed by
the HXMT ME and ACD (black), as well as their FFT filtered profiles
(green). Panels~(b), (e): detrended light curve. Panels~(c), (f):
Morlet wavelet power spectra. The magenta lines outline the
significance level of 99.9\%. \label{wv_h}}
\end{figure}

\begin{figure}
\centering
\includegraphics[width=\linewidth,clip=]{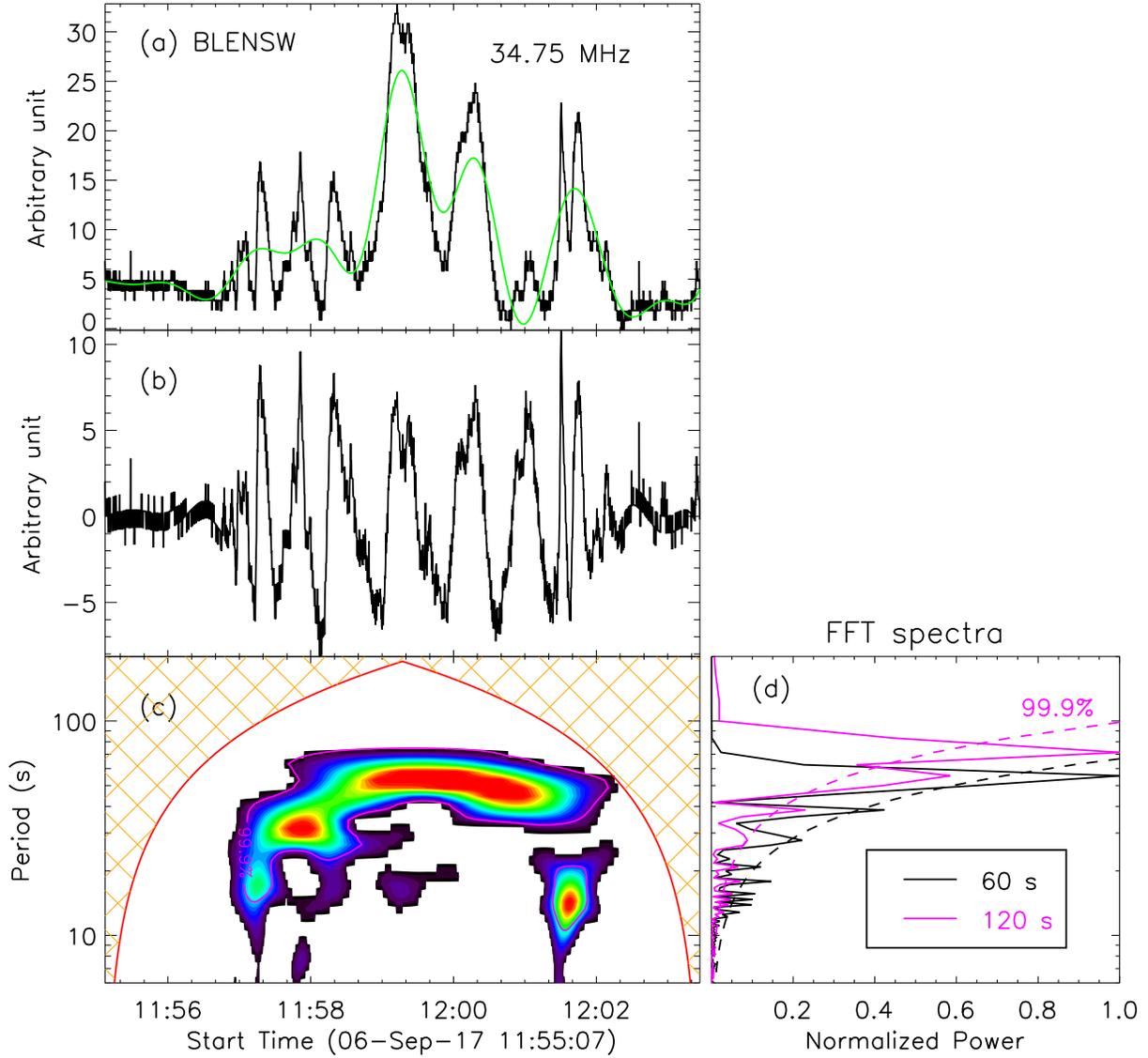}
\caption{Panel~(a): Full-disk light curves (black) observed by the
BLENSW at a frequency of 34.75~MHz (black), and its FFT filtered
profile (green). Panels~(b): detrended light curve. Panels~(c):
Morlet wavelet power spectra. The magenta lines outline the
significance level of 99.9\%. Panel~(d): FFT power spectra of the
detrended light curves with cutoff periods of 60~s (black) and 120~s
(magenta), respectively. While their significance levels are marked
by the dashed lines. \label{wv_r}}
\end{figure}

The flare light curves at wavelengths of MUV Balmer continuum,
Ly$\alpha$, and HXR are characterized by a number of small
pulsations, referred to as QPPs. Similar to previous observations
\citep{Ning14,Li17,Milligan17,Feng20}, these small pulsations
superimposed on the strong background. To look more closely at the
periodicity of this flare, we apply the wavelet transform method
\citep{Torrence98} to the detrended light curves, where detrending
is obtained by filtering out the longest periods (i.e., the
long-term trend) from the original/raw time series. Here, the
wavelet mother function of `Morlet' is used for the wavelet
analysis. The results are shown in Figures~\ref{wv_l}, \ref{wv_h}
and \ref{wv_r}. Based on the fast Fourier transform (FFT) method,
the raw light curves measured by LYRA, Insight-HXMT, and BLENSW are
detrended using a cutoff threshold of 60~s, thereby enhancing the
periods that are shorter than 60~s. Thus, the periodic features with
short periods can be highlighted in the wavelet power spectrum
\citep{Feng17,Ning17,Milligan17}.

Figure~\ref{wv_l} presents the Morlet wavelet analysis results for
the LYRA data from two channels. The top two panels draw the raw
light curves of Ly$\alpha$ (channel~1) and MUV Balmer continuum
(channel~2) emissions, the long-term trend obtained by FFT-filtering
are overplotted in green. Two main pulses appear in the raw light
curves during about 11:55:40$-$11:58:20~UT, as indicated by the two
green arrows in panels~(a) \& (d). The oscillatory amplitude of the
two main pulses is much larger than that of the short-period QPPs
studied in this study, which may result in a weak power of the
wavelet analysis. Therefore, to suppress the long-period trend that
is caused by the two main pulses, a cutoff period of $\sim$60~s is
used to the raw light curves. The detrended light curves are
obtained by subtracting the FFT filtered time series, as plotted in
the middle panels. They both show a series of pulsations, and each
pulsation has an anharmonic and symmetric triangular profile shape,
which can be regarded as the signature of non-stationary QPPs. The
bottom panels plot the wavelet power spectra for the detrended light
curves in Ly$\alpha$ (c) and MUV Balmer continuum (f) emissions,
both of which show an enhanced power over a wide range of periods
during the impulsive phase of the powerful flare, i.e., between
about 11:56~UT and 12:02~UT, implying the multi-periodicity QPPs.
The bulk of the detections in the two power spectra are evident at
periods of about 20$-$55~s. Interestingly, the quasi-periods appear
to have different lifetimes. For instance, the quasi-periods of
$\sim$20$-$30~s are seen to happen from about 11:56~UT to 11:57~UT
only, while the 30$-$50~s QPPs are seen to occur between about
11:56~UT and 11:59~UT, and the quasi-periods of 50$-$55~s are seen
to appear during $\sim$11:57$-$12:02~UT. However, the 50$-$55~s QPP
signal is very close to the cutoff threshold, which might be an
artefact of detrending. Similarly to previous studies
\citep{Kupriyanova10,Milligan17}, we then extracted the detrended
time series from the raw LYRA data with a long cutoff threshold of
120~s, as shown in Figure~\ref{wv_l2}. The wavelet power spectra
show a much broader range of periods, such as 20$-$100~s. The
strongest power appears at the periods of $\sim$60$-$100~s between
$\sim$11:55:40~UT and $\sim$11:58:20~UT, which is largely due to the
two main pulses shown by the green arrows in Figure~\ref{wv_l}~(a)
and (d). They are not suppressed when we used a cutoff threshold of
120 s for detrending, as shown in Figure~\ref{wv_l2}~(a) \& (b). On
the other hand, the short periods of about 20$-$55~s exhibits a
little weak power, but they are still inside the 99.9\% significance
level, which confirms that the short periods in LYRA channels~1 and
2 are not artifacts of the detrending process.

\begin{figure}
\centering
\includegraphics[width=\linewidth,clip=]{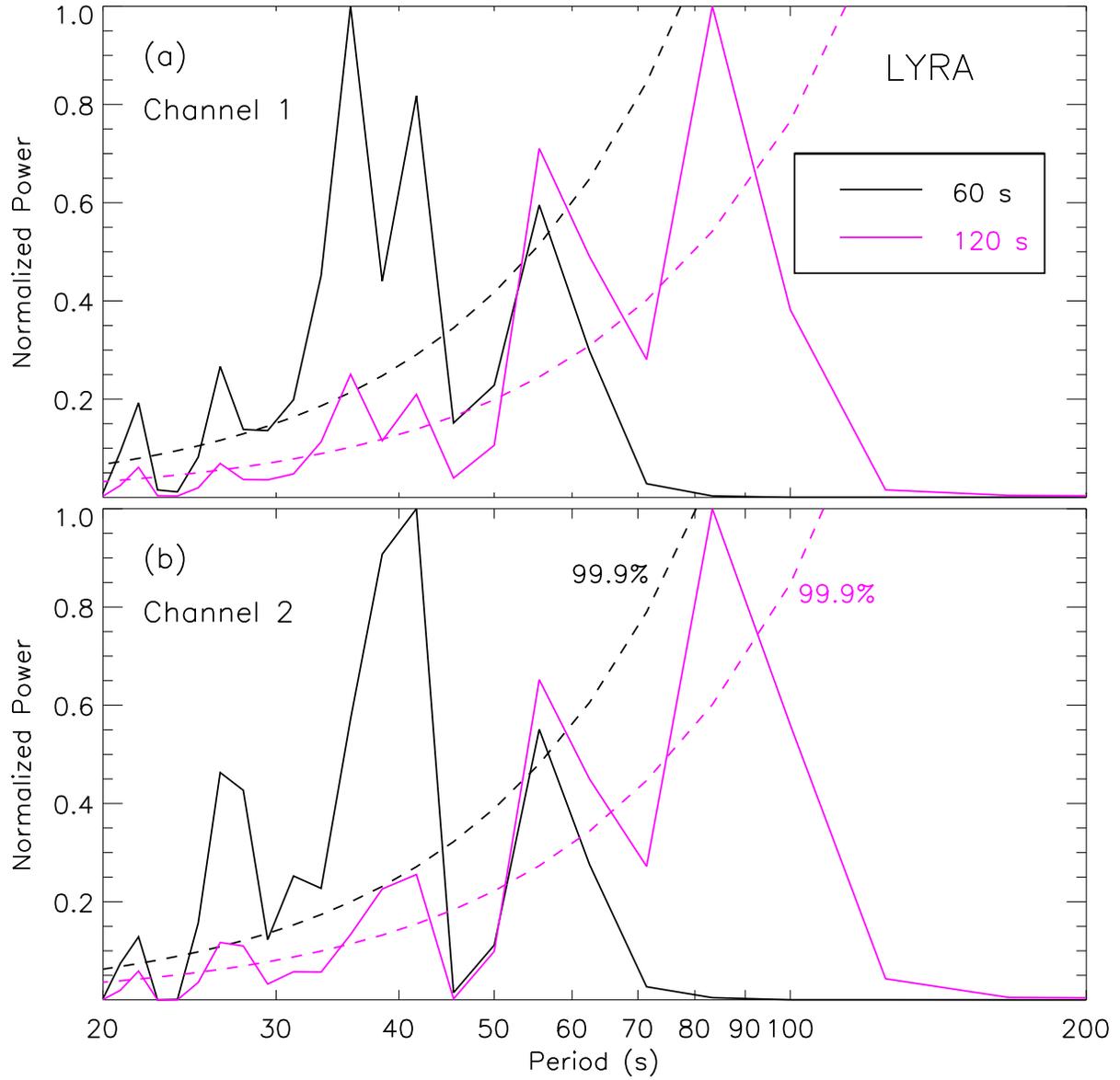}
\caption{FFT power spectra derived from the LYRA at channels~1 (a)
and 2 (b). The detrended light curves obtained from the cutoff
periods at 60~s (black) and 120~s (magenta), respectively. Their
significance levels are marked by the dashed lines. \label{ffts}}
\end{figure}

The wavelet power spectra in Figure~\ref{wv_l2} show a broad range
of periods with a brightest core and weaker borders. Using the
Lomb-Scargle periodogram method \citep{Scargle82}, the periodogram
analysis is performed to the detrended light curves.
Figure~\ref{ffts} presents the normalized FFT power spectra at LYRA
channels~1 (a) and 2 (b), and two different values of cutoff
thresholds are used for detrending, e.g., 60~s (black) and 120~s
(magenta), respectively. The dashed lines in the FFT spectra
represent the significance levels of 99.9\%, which are estimated
from the red noise in quasi-periodic signals
\citep{Vaughan05,Liang20,Anfinogentov21}. The short periods of
$\sim$20$-$55~s can be simultaneously seen in all the FFT power
spectra, but the quasi-periods of $\sim$20$-$40~s are quite weak in
the FFT power spectra from the detrending light curves by applying a
big cutoff threshold of 120~s, as shown by the magenta lines. On the
other hand, the long period of roughly 80~s is only found in the FFT
power spectra from the detrending light curves by applying the 120-s
cutoff threshold, and it is well suppressed by the 60-s cutoff
threshold, which is similar to Morlet wavelet analysis results. The
FFT spectra further confirm that the short periods of
$\sim$20$-$55~s are not artifacts of the detrending process.
However, it is impossible to determine the duration of the short
periods, especially the lifetime of periods between 50$-$55~s, since
they are also seen at the border edges of the strongest power in the
wavelet power spectra in Figure~\ref{wv_l2}. We also notice that the
80-s period has been studied in the flare emission at wavelengths of
Ly$\alpha$ \citep{Li20d} and HXR emission \citep{Zhang21},
respectively. Therefore, we only focused on the short periods of
$\sim$20$-$55~s in this study.

Figure~\ref{wv_h} shows the Morlet wavelet analysis results for the
Insight-HXMT data at ME and ACD channels. Using the same FFT method,
the long-term trend and detrended time series are separated from the
raw light curves, as shown in the top and middle panels. The two
main pulses are also found in the raw light curves at HXR channels
(a \& f), which are similar to the double main pulses measured by
the LYRA from channels 1 and 2. Then the wavelet analysis technique
is applied to the detrended time series, as shown in the panels~(e)
and (f). Thus, the quasi-periods in flare X-ray emissions can be
determined in their wavelet power spectra. They both show an
enhanced power over a broad range of periods at roughly 20$-$55~s
during the flare impulsive phase, suggesting that they are multiple
periods, which is similar to that in Ly$\alpha$ and MUV Balmer
continuum emissions. Moreover, the multiple periods appear to show
different lifetimes. The quasi-periods of roughly 20$-$40~s are
found to appear between $\sim$11:56~UT and $\sim$11:58~UT, and they
could be co-existence. While the periods of about 50$-$55~s are seen
to occur from about 11:57~UT to 12:01~UT. That is, they are
existence of multiple oscillatory signals with periods of about
20$-$55~s and have different lifetimes.

Figure~\ref{wv_r}~(a)$-$(c) presents the Morlet wavelet analysis
results for the BLENSW radio data at the low frequency of 34.75~MHz.
Panels~(g) \& (h) plot the light curve and its detrended light curve
by applying a cutoff threshold of 60~s. They both show a number of
repeated and triangular peaks from about 11:57~UT to 12:02~UT, which
could be regarded as the flare QPPs. Moreover, these peaks show a
well one-to-one correspondence between the raw and detrended light
curves, indicating that they are really QPP signals rather than the
artefact of detrending. Panel~(c) draws the wavelet power spectrum,
which clearly shows an enhanced power over a broad range, suggesting
multiple periods in the low-frequency radio flux. On the other hand,
the flare QPPs with multiple periods starts at about 11:57~UT, which
are later than those begin at HXR channels, such as 1-minute time
delay. Panel~(d) further shows the normalized FFT power spectra of
the low-frequency radio flux by applying the cutoff thresholds of
60~s (black) and 120~s (magenta) for detrending. Similar to previous
LYRA results in Figure~\ref{ffts}, the short periods of
$\sim$20$-$55~s can be simultaneously seen in these two FFT power
spectra, and they are corresponding well with each other, although
their peaks obtained from the 120-s cutoff threshold are a bit low.
This confirms that the short periods are not artifacts of the
detrending process.

\begin{figure}
\centering
\includegraphics[width=\linewidth,clip=]{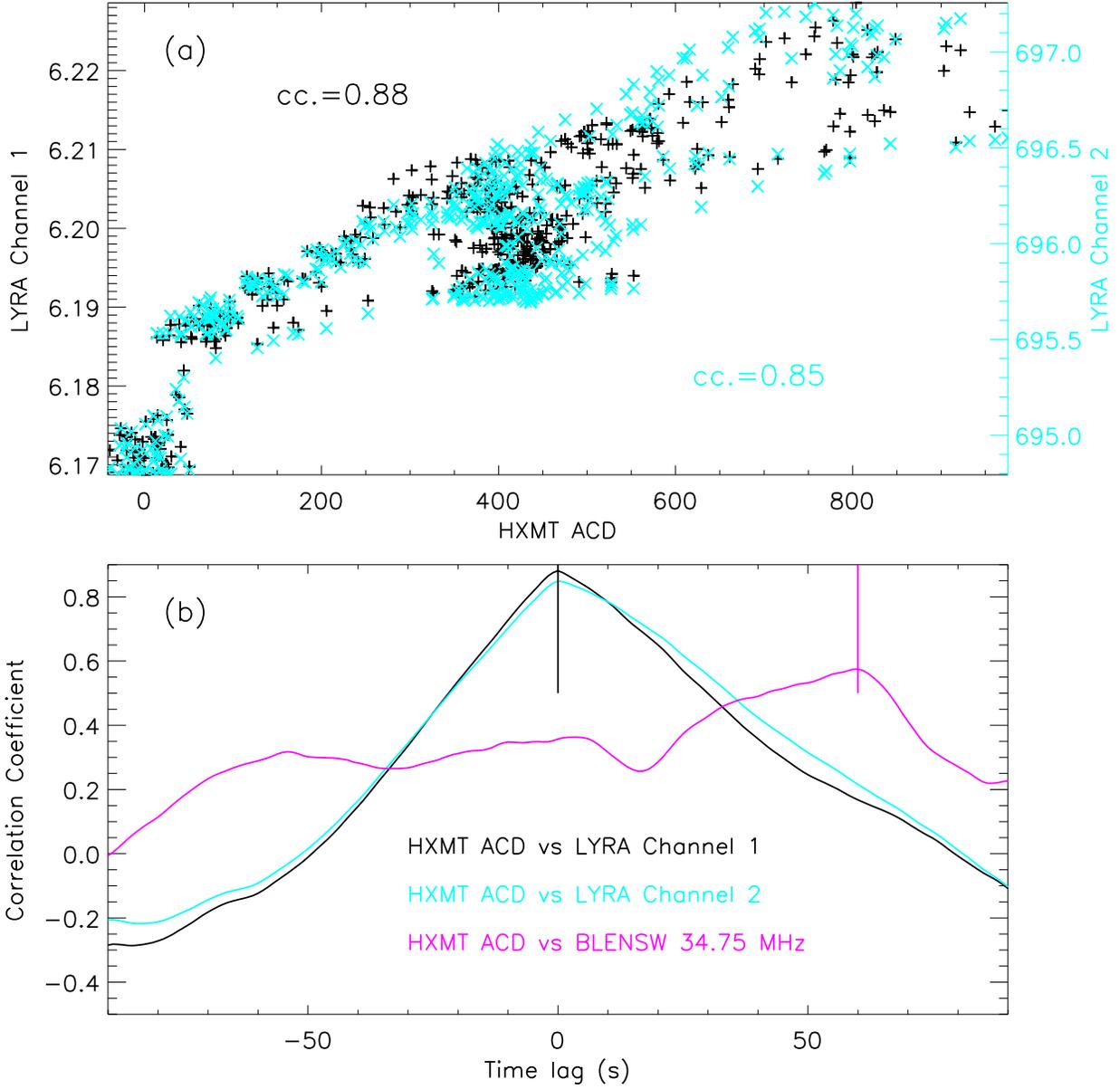}
\caption{Pane~(a): Ly$\alpha$ and MUV Balmer continuum emissions
depend on the HXR flux during the impulsive phase of solar flare,
i.e., between $\sim$11:55~UT and $\sim$12:03~UT. The correlation
coefficients (cc.) are given. Panel~(b): Cross-correlation analysis
results, which respect to the raw light curves at HXMT ACD.
\label{tlag}}
\end{figure}

Similar non-stationary QPPs with multiple periods can be
simultaneously detected at wavelengths of Ly$\alpha$, MUV Balmer
continuum, HXR and low-frequency radio during the impulsive phase of
a powerful solar flare. To further study their relationship, we then
draw the correlation between the HXR flux and Ly$\alpha$ emission
(black), as well as the MUV Balmer continuum emission (cyan), as
shown in Figure~\ref{tlag}~(a). Here, the LYRA data has been
interpolated into a time cadence of 1~s, which is same to that of
the Insight-HXMT data, so that the correlation between two different
instruments can be well underlined. Notice that the raw light curve
rather than the detrended one is used here. The high correlation
coefficients (cc.) are obtained between them, which are of 0.88 and
0.85, respectively. On the other hand, low correlation coefficients
are found between the SXR flux at GOES~1$-$8~{\AA} and the MUV
Balmer continuum or Ly$\alpha$ emissions, i.e., coefficients of 0.12
or 0.19. These correlation coefficients further confirm that both
the Ly$\alpha$ and MUV Balmer continuum emission recorded by the
LYRA at channels~1 and 2 show nonthermal temporal behaviors rather
than the thermal profiles, which agrees well with previous findings
\citep[e.g.,][]{Dominique18,Milligan20}.

The time delay is found between the HXR and low-frequency radio
data, as shown in Figure~\ref{flux}~(b). To further investigate
their links, we then perform their cross-correlation analysis, as
well as the cross-correlation analysis between the HXR and LYRA
data, as shown in Figure~\ref{tlag}~(b). A maximum correlation
coefficient of $\sim$0.57 between the HXR and low-frequency radio
data is seen at the time lag of around 60~s, as indicated by the
magenta vertical line, which indicates a time delay of about 60~s
between them. On the other hand, the maximum correlation
coefficients of 0.88 and 0.85 are found between the HXR and LYRA
data at the time lag of 0~s (black vertical line), suggesting no
time delay between them, which is similar to previous findings in
Figure~\ref{flux}. We would like to state that the raw light curves
rather than the detrended time series are used for the
cross-correlation analysis. The maximum correlation coefficient
between the HXR and radio data is only $\sim$0.57, which is slightly
lower than those between HXR and LYRA data. This is because that the
pulses in radio flux show large amplitudes, for instance, increasing
and decaying rapidly with respect to their background emission,
while the HXR/LYRA light curves only exhibit small-amplitude
pulsations, as shown in Figure~\ref{flux}~(b). The positive
correlation between the HXR and radio signals suggests that they are
produced by the same process of energy releases, i.e., magnetic
reconnection during the solar flare.

\section{Summary and Discussion}
Using the observations measured by the PROBA2/LYRA, the
Insight-HXMT, and the BLENSW, we investigate the non-stationary QPPs
with multiple periods during the impulsive phase of the X9.3 flare
on 2017 September 06, which was the most powerful flare of solar
cycle 24. Based on the wavelet analysis technique \citep{Torrence98}
and the Lomb-Scargle periodogram method \citep{Scargle82}, the
multiple periods from roughly 20~s to about 55~s are simultaneously
identified during the flare impulsive phase in the Ly$\alpha$
emission, the MUV wavelengths around 2000~{\AA}, as well as HXR and
radio channels. Multi-mode QPPs with non-stationary properties were
studied in the microwave/HXR emission during solar flares
\citep{Inglis09,Kolotkov15}. Using the wavelet analysis method,
\cite{Inglis09} demonstrated that the multiple periods of multi-mode
QPPs could be co-existed nearly simultaneously, and there was almost
not significant period (or frequency) shift over time. In this
study, the flare QPPs are detected simultaneously in multiple
wavelengths, i.e., MUV, Ly$\alpha$, HXR, and low-frequency radio. On
the other hand, the multiple periods observed in this flare are not
co-existing. The quasi-periods of 50$-$55~s are affected by the two
main pulses, which makes it difficult to determine their onset time
and lifetimes. However, the quasi-periods of $\sim$20$-$40~s can
only be found during $\sim$11:56$-$11:59~UT, which are obviously
shorter than the 50$-$55-s periods. This suggests that the multiple
periods have different lifetimes. Thus, the flare QPPs observed here
are regarded as multiple periods with different lifetimes. Previous
studies suggested that the MUV wavelengths around 2000~{\AA} agreed
well with the hydrogen Balmer continuum emission produced in the
optically thin chromosphere \citep{Dominique18}. Therefore, this is
the first report of flare-related QPPs with multiple periods in the
MUV Balmer continuum emission.

The QPP behaviors of the X9.3 flare has been studied at wavelengths
of radio, SXR, HXR, and even $\gamma$-ray
\citep{Kolotkov18,Karlicky20,Li20a,Zhang21}. \cite{Kolotkov18} first
investigated the QPP signals during the X9.3 flare. They found that
the QPP periods varied from around 12~s to 25 s in the thermal
emission during the flare impulsive and decay phases, and attributed
them to be the sausage oscillations of flaring loops. Next,
\cite{Li20a} studied the flare-related QPPs with periods of about
20$-$30~s at channels of radio, HXR and $\gamma$-ray during the
impulsive phase, and the similar periods were also reported by
\cite{Zhang21}. Then, \cite{Karlicky20} detected multiple periods
(i.e., 1$-$2~s, 5.3$-$8.5~s and 11$-$30~s) mainly in the radio
emission during the pre-impulsive and impulsive phases of the X9.3
flare. In this study, we report the multi-periodic QPPs during the
flare impulsive phase in Ly$\alpha$, MUV Balmer continuum,
low-frequency radio and HXR channels simultaneously. The short
periods of 20$-$40~s are detected during $\sim$11:56$-$11:59~UT,
which are similar to previous findings in the radio, HXR and
$\gamma$-ray channels \citep{Li20a,Zhang21}. However, they could not
study the QPP behaviors after 11:59~UT, largely due to their
observational limitations. Here, the long periods of $\sim$50$-$55~s
are also found from roughly 11:57~UT to 12:02~UT. We would
like to point out that a much longer period of roughly 80~s could
also be seen in the LYRA data, and it is mostly caused by the two
main pulses during $\sim$11:55:40$-$11:58:20~UT in the raw light
curves, which are similar to the intermittent feature in the HXR
emission \citep{Zhang21}. However, it is not suitable for a typical
QPP, since it only remains two cycles (e.g., two pulses). Moreover,
the flare QPPs with periods of about 60$-$80~s have been reported in
the Ly$\alpha$ emission during solar flares \citep{Li20d}. So, only
the short periods between $\sim$20~s and $\sim$55~s are studied at
wavelengths of Ly$\alpha$, MUV Balmer continuum, HXR and
low-frequency radio in this study.

It is necessary to discuss the generation mechanism that can be
responsible for the multi-periodic QPPs detected simultaneously at
wavelengths of Ly$\alpha$, MUV Balmer continuum, HXR, and
low-frequency radio. The Ly$\alpha$ irradiance during solar flare is
found to be closely related with the MUV Balmer continuum emission,
and it also shows highly synchronous with the HXR emission
\citep{Rubio09,Dominique18}. On the other hand, the flare radiation
in HXR and low-frequency radio channels during the impulsive phase
is generally produced by the bi-directional nonthermal electrons
accelerated by the magnetic reconnection \citep[e.g.,][]{Benz17}.
Therefore, the flare QPPs simultaneously observed in the Ly$\alpha$,
MUV Balmer continuum, HXR and low-frequency radio channels are most
likely to be triggered by a same repeated energy release process,
i.e., the periodic regime of magnetic reconnection
\citep{Li17,Li20a,Clarke21}. The pulsed, bi-directional electron
beams are accelerated by the intermittent magnetic reconnection
during the flare impulsive phase. The downward accelerated electrons
precipitated toward the chromosphere along the flare loop, resulting
in quasi-periodic enhancements in Ly$\alpha$, MUV Balmer continuum
and HXR emissions, while the upward electron beams escaped along an
open magnetic field line and generated radio pulsations at the low
frequency of 34.75~MHz. The time delay of about 60~s between the HXR
QPP and the low-frequency radio QPP is attributed to the traveling
time of the upward electron beams propagating outward from the Sun
\citep{Li15,Clarke21}. It is generally accepted that the type III
radio burst is produced by the electron beam traveling along the
magnetic field and outwards from the Sun \citep{Reid14}. In
particular, the radio emissions at frequencies of about
200$-$0.03~MHz are often dominated by the plasma emission mechanism
\citep[e.g.,][]{Gary89}. That is, the radio frequency ($f$) is
roughly proportional to the local electron density ($n_e$), such as
$f\thickapprox8980\sqrt{n_e}$ \citep[e.g,.][]{Lu17}. Based on the
electron density model developed by \cite{Vrsnak04}, the radio flux
at the frequency of 34.75~MHz is produced at the heliocentric height
of roughly 2.5R$_{\odot}$ (R$_{\odot}$ represents the solar radius)
above the Sun, and it agrees with previous models
\citep[e.g.,][]{Gary89,Krupar14}. Then, the average speed of the
electron beam is estimated to be $\sim$0.1$c$ ($c$ is the light
speed), which is consistent with previous findings in the range of
0.1$c-$0.5$c$ \citep{Dulk87,Reid14,Clarke21}. Considering the
multiple periods have different lifetimes, the repetitive magnetic
reconnection is probably to be induced rather than spontaneous. The
similar short periods have been found in the major X9.3 flare
\citep{Kolotkov18} or in the flare Ly$\alpha$ emission
\citep{Van11}, which were interpreted in terms of sausage
oscillations. Therefore, the multi-periods of $\sim$20$-$55~s might
be modulated by the sausage oscillations of hot plasma loops in the
flare region \citep{Chen15,Guo16}. The flare-related QPPs with
multiple periods could also be related to the length of flare loops
and each other \citep[e.g.,][]{Reznikova11,Pugh19}. However, the
other generation mechanisms, for instance, driven directly by MHD
waves, or by LRC-circuit oscillations, could not be completely ruled
out, largely due to the absence of the spatially resolved
information, i.e., high-resolution images at wavelengths of MUV,
Ly$\alpha$, radio, and HXR.


It should be stated that the Insight-HXMT is primarily used to scan
the Galactic Plane and study X-ray binaries and gamma-ray bursts
\citep{Zhang20}. However, the present study reveals that it can also
capture the HXR emission of solar flares, like, in this case, the
X9.3 flare on 2017 September 06. The X9.3 flare was measured by the
Konus-Wind in HXR and $\gamma$-ray wavelengths, but these
observations only covered a total duration of about 250~s
\citep{Li20a}, which makes it impossible to study the flare-related
QPPs during the whole impulsive phase. Here, using the observations
recorded by the Insight-HXMT, the PROBA2/LYRA and the BLENSW, we
detect the similar flare-related QPP behaviors at channels of HXR,
Ly$\alpha$, MUV Balmer continuum emission, and the low-frequency
radio, for instance, the multi-periods with different lifetimes. Our
findings confirm the nonthermal temporal behaviors of flare
radiation in Ly$\alpha$ and MUV Balmer continuum, which is
consistent with previous observational results. In the study of
\cite{Dominique18}, the derivative of the GOES SXR flux was used as
a proxy for the HXR light curve, largely due to the absence of HXR
measurements from solar telescopes. Thus, this study also provides
an idea that some astronomical satellites such as Insight-HXMT or
Fermi could also be used to study solar powerful eruptions, in
particular the larger solar flare.

\acknowledgments The authors thank the anonymous referee for his/her
valuable comments and suggestions. LYRA is a project of the Centre
Spatial de Li\`{e}ge, the Physikalisch-Meteorologisches
Observatorium Davos and the Royal Observatory of Belgium funded by
the Belgian Federal Science Policy Office (BELSPO) and by the Swiss
Bundesamt f\"{u}r Bildung und Wissenschaft. This work also made use
of the data from the Insight-HXMT mission, a project funded by China
National Space Administration (CNSA) and the CAS. This work is
supported by NSFC under grants 11973092, 12073081, 11921003,
11873095, 11790302, U1931138, U1731241, U1838202, U1838201,
U1938102, U1838110, as well as CAS Strategic Pioneer Program on
Space Science, Grant No. XDA15052200, and XDA15320301. D.~Li is
supported by the Specialized Research Fund for State Key
Laboratories. The Laboratory No. 2010DP173032. The Insight-HXMT team
gratefully acknowledges the support from the National Program on Key
Research and Development Project (Grant No. 2016YFA0400800) from the
Minister of Science and Technology of China (MOST) and the Strategic
Priority Research Program of the Chinese Academy of Sciences (Grant
No. XDB23040400). M. Dominique thanks the European Space Agency
(ESA) and BELSPO for their support in the framework of the PRODEX
Programme.

\begin{table}
\caption{The details of observational instruments in this study.}
\centering
\begin{tabular}{c c c c c c c}
\hline
Instruments  &  Channels  &  Cadence   &    Wavelengths      &  Bandpass   \\
\hline
             & Channel 1  &  0.05~s    &  1200$-$1230~{\AA}  &  Ly$\alpha$ \\
LYRA         & Channel 2  &  0.05~s    &  1900$-$2220~{\AA}  &  MUV        \\
             & Channel 4  &  0.05~s    &   1$-$200~{\AA}     &  SXR/EUV    \\
\hline
GOES-16      &   XRS      &   1~s      &    1$-$8~{\AA}      &  SXR        \\
\hline
             &   ME       &   1~s      &    5$-$30~keV       &  SXR/HXR    \\
HXMT         &   ACD      &   1~s      &    $>$100~keV       &  HXR        \\
\hline
BLENSW       &           &  0.25~s    &  $\sim$20$-$76~MHz   &  Radio        \\
\hline
\end{tabular}
\label{tab1}
\end{table}

\clearpage
\appendix
\renewcommand\thefigure{\Alph{figure}}
\section{Appendix}
\setcounter{figure}{0}

\begin{figure}
\centering
\includegraphics[width=\linewidth,clip=]{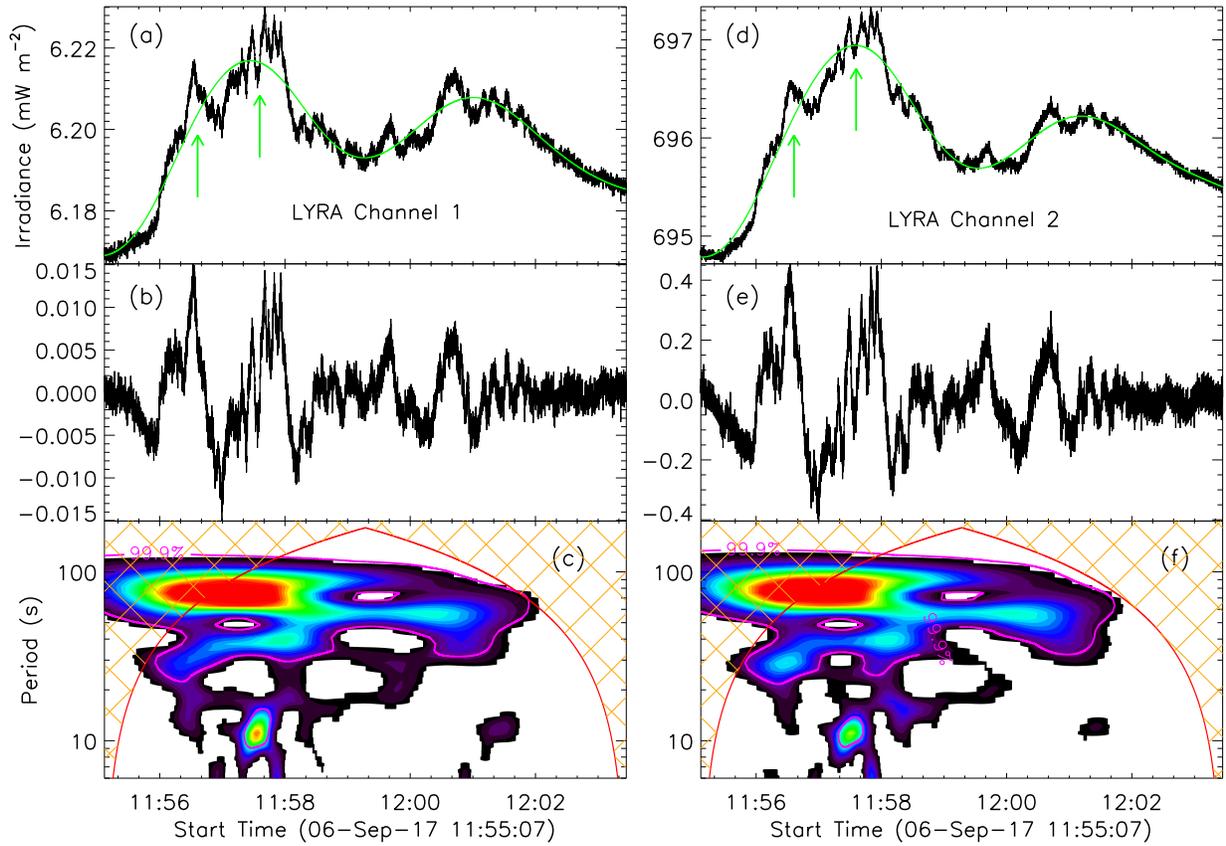}
\caption{Following in Figure~\ref{wv_l}, the Morlet wavelet analysis
results for the LYRA data. Here, the detrending signals and their
wavelet power spectra are obtained from the cutoff period of 120~s.
The magenta lines outline the significance level of 99.9\%.
\label{wv_l2}}
\end{figure}

\end{document}